\documentclass[a4paper,10pt]{article}
\usepackage[margin=1.0in]{geometry}

\usepackage{amssymb,amsfonts,amsmath,epsfig,float,graphicx}

\usepackage {url}

\def\upd{{\rm d}}
\newcommand{\be}{\begin{eqnarray}}
\newcommand{\ee}{\end{eqnarray}}

\def\Io{{\mathbb I}}

\def\Ro{{\mathbb R}}

\def\eev{{\bf e}}

\def\kk{{\bf k}}

\def\xx{{\bf x}}

\def\kkph{{\kk^{\mbox{\tiny ph}}}}
\def\kph{k^{\mbox{\tiny ph}}}

\def\uc{q_{\mbox{\tiny el}}}

\def\ranglestar{\rangle_{\hskip -2pt _*}}
\def\ranglec{\rangle^{\mbox{\tiny c}}}

\renewcommand{\Re}{\,{\rm Re}\,}

\begin{document}
\title{Reducible Quantum Electrodynamics.
I. The Quantum Dimension of the 
Electromagnetic Field}
\author{Jan Naudts}

\maketitle

\begin{abstract}
In absence of currents and charges the quantized electromagnetic field can be 
described by wave functions which for each individual wave vector are normalized 
to one.
The resulting formalism involves reducible representations of the Canonical 
Commutation Relations. The corresponding paradigm is a space-time filled with 
two-dimensional quantum harmonic oscillators. Mathematically, this is equivalent 
with two additional dimensions penetrated by the electromagnetic waves.
  
 \end{abstract}

\section*{Preface}

It is common to consider classical mechanics as a limiting case of quantum mechanics.
This is the correspondence principle of Niels Bohr.
The limit is usually taken by letting Planck's constant $\hbar$ tend to zero.
The present paper starts from a different paradigm. Quantum mechanics is considered
to happen in a direction orthogonal to classical space-time.
This is what I call the {\sl quantum dimension.} This point of view leads to a
reformulation of standard quantum electrodynamics.
The present paper treats the free photon part of it.

The most visible difference between the present formalism and the standard one is that
integrations in momentum space disappear as part of the field operators and move towards the
evaluation of expectation values. Indeed, any function
of the wave vector $\kk$ becomes an operator itself and works on
wave functions indexed by a wave vector.
In the standard formalism there is one annihilation operator $a_\kk$ for each wave vector $\kk$.
Here, the annihilation operator can be thought of as belonging to
a single quantum harmonic oscillator with a potential $V(x^4)$
which depends on one additional spatial coordinate $x^4$.
The motion in this direction is purely quantal
and is orthogonal to the 4 directions of space-time.
Because the electromagnetic waves have two orthogonal polarizations
two additional dimensions are needed instead of one.

The idea of additional space-time dimensions is of course not new.
It started with the fifth dimension of Kaluza-Klein and received continued interest.
See for instance \cite{OW97, WP06}.
Related to the present vision
is the idea that space-time is a 4-dimensional membrane living in a 6-dimensional space
\cite {RS83,VM85,GW87}.
Here the emphasis is on the quantum nature of the two additional dimensions.

The present reformulation of quantum field theory finds its origin in a series of papers
by Czachor and collaborators in a study of reducible representations of the
canonical commutation relations --- see \cite{CM00,CN06,CW09} and references given in these
papers.
In particular the idea to turn the frequency of the harmonic oscillator into
an operator is borrowed from these works.
An important difference between the present formalism and
that of \cite{CW09} is that here is chosen for a vacuum
which is unique and structureless.
This is justified because quantum uncertainty leads anyhow
to fluctuations around the vacuum state.

Further input to the present work comes from coherent state quantization.
See for instance Part II of \cite{GJP09}.
Note however that here it is not the intention to quantize the classical world
but rather to extend the classical description of
an electromagnetic wave into the quantum world.

Finally, the present work has been influenced by
recent experiments in quantum optics \cite {SZ97},
such as those of \cite{HOM87}, \cite{GBDSGKBRS07,BB11} and \cite {KBRSMSS11}.
In my opinion the present formalism is more suited than
standard Quantum Electrodynamics (QED) for a detailed description of these experiments.
Forthcoming research has to validate this statement.

Why study reducible representations of canonical commutation relations?
One argument in favor of irreducible representations, found in the literature \cite{BLT75},
is to exclude classical mechanics. Another argument is that reducible representations can be
decomposed as integrals over irreducible representations.
The latter argument can be applied to classical mechanics as well.
However, limiting the study of classical mechanics to its irreducible representations
would cripple the theory. 
A further argument is the intuition that mechanical theories are stabilized by deforming them
whenever possible. 
In this spirit Weyl's formalism (see for instance \cite{MJE49}, or \cite{HRS08} for a recent paper)
describes quantum mechanics as a deformation of classical mechanics.
By the deformation the theory becomes less reducible.

The arguments given above in favor of irreducible representations
are rather vague, while the reducible representations
have one very concrete advantage: they allow to avoid some of the intrinsic difficulties of
the standard theory. The derivation of QED found in standard textbooks
is far from obvious and is not built on mathematically sound principles. More rigorous
approaches, such as the one described in \cite {HK64} or \cite{BLT75}, cannot justify the daily practices
of researchers in quantum optics or in elementary particle physics.

The next section develops the formalism in the context of scalar bosons.
Section 2 deals with free electromagnetic fields.
Some conclusions are drawn in the final section.
A few calculations are transferred to the Appendices.

\section{The scalar boson field}
\label {section:scalar}

\subsection{The classical wave equation}

\def\Ecl{E_{\mbox{\tiny cl}}}
\def\ecl{e_{\mbox{\tiny cl}}}
\def\lu{{l}}

The general real solution of the wave equation $\square\phi=0$
can be written in terms of an arbitrary complex profile function $f(\kk)$ as 
\be
\phi(x)&=&\int_{\Ro^3}\upd\kk\,\frac{\lu}{N(\kk)}
\left[f(\kk)e^{i(\kk\cdot\xx-c|\kk|t)}
+\overline{f(-\kk)}e^{i(\kk\cdot\xx+c|\kk| t)}\right]\cr
&=&
2\Re \int_{\Ro^3}\upd\kk \,\frac{\lu}{N(\kk)}
f(\kk)e^{-ik_\mu x^\mu},
\label{weq:exp}
\ee
where $k_0=|\kk|$.
The constant length $\lu$ has been added to make the field $\phi(x)$ dimensionless.
The so-called normalization factor $N(\kk)$ is the usual one
\be
N(\kk)=\sqrt{(2\pi)^32|\kk|}.
\label{scalar:norm}
\ee
The choice of normalization leads further on to a satisfactory physical interpretation
of the profile function $f(\kk)$.

The total energy is given by
\be
\Ecl
&=&\frac {\hbar c}{2\lu^2}\int_{\Ro^3}\upd\xx\,\left[\left(\frac{\partial\phi}{\partial x^0}\right)^2
+\sum_\alpha\left(\frac{\partial\phi}{\partial x^\alpha}\right)^2
\right].
\label{scalar:classenerg}
\ee
It is a conserved quantity. Indeed, one has
\be
\frac{\upd \Ecl}{\upd x^0}
&=&\frac{\hbar c}{\lu^2}\int_{\Ro^3}\upd\xx\,\left[\left(\frac{\partial\phi}{\partial x^0}\right)
\left(\frac{\partial^2\phi}{\partial (x^0)^2}\right)
+\sum_\alpha\left(\frac{\partial\phi}{\partial x^\alpha}\right)
\left(\frac{\partial\,}{\partial x^\alpha}\frac{\partial\phi}{\partial x^0}\right)
\right]\cr
&=&0.
\ee
Use partial integration and the wave equation to obtain the above result.
From the plane wave expansion (\ref{weq:exp}), using the wave equation, one obtains
\be
\Ecl=\int_{\Ro^3}\upd\kk \,\hbar c|\kk||f(\kk)|^2.
\label{weq:classen}
\ee
The interpretation in the context of quantum mechanics is standard.
The factor $|f(\kk)|^2$ is the density of particles with wave vector $\kk$
and corresponding energy $\hbar c|\kk|$.

\subsection{Coherent states}

The coherent states \cite{GJP09} of a quantum harmonic oscillator
are closest to the states of classical
mechanics because they minimize Heisenberg's uncertainty relations.
They are eigenstates of the annihilation operator $\hat a$ with complex eigenvalue $z$
\be
\hat a|z\ranglec=z|z\ranglec.
\ee
The Hamiltonian of the quantum harmonic oscillator reads
\be
\hat H=\hbar c|\kk| a^\dagger a.
\label{scal:ham}
\ee
Given a coherent state $|F(\kk)\ranglec$ the expectation of the quantum energy reads
\be
\langle F(\kk)|\hat H|F(\kk)\ranglec
&=&\hbar c|\kk||F(\kk)|^2.
\ee
Take now $F(\kk)$ equal to
\be
F(\kk)&=&\lu^{-3/2}f(\kk),
\label{coh:Ff}
\ee
where $\lu$ is the constant introduced before. It makes $F(\kk)$ dimensionless.
Then (\ref {weq:classen}) can be written as
\be
\Ecl&=&\lu^3\int\upd\kk\,\langle F(\kk)|\hat H|F(\kk)\ranglec.
\label{quant:energ}
\ee

For each value of $\kk$ the symbol $|F(\kk)\ranglec$ is a coherent state.
This means that one harmonic oscillator is associated with
each value of the wave vector $\kk$.
The standard interpretation is that the space is filled
with harmonic oscillators.
This interpretation is taken literally further on.

\subsection{Field operators}

From (\ref {weq:exp}) follows that the classical field can be written as
\be
\phi(x)&=&\int\upd\kk\,\frac{\lu^{5/2}}{N(\kk)}\langle F(\kk)|\hat\phi(x)|F(\kk)\ranglec,
\ee
with a field operator $\hat\phi(x)$ defined by
\be
\hat\phi(x)=e^{-ik_\mu x^\mu} \hat a +  e^{ik_\mu x^\mu} \hat a^\dagger.
\ee
A short calculation gives
\be
[\hat\phi(x),\hat H]&=&e^{-ik_\mu x^\mu}
\hbar c|\kk| [\hat a,\hat a^\dagger \hat a]
+ e^{ik_\mu x^\mu} \hbar c|\kk|[\hat a^\dagger,\hat a^\dagger \hat a]\cr
&=& e^{-ik_\mu x^\mu}\hbar c|\kk| a- e^{ik_\mu x^\mu}\hbar c|\kk|\hat a^\dagger\cr
&=&i\hbar c\frac{\partial\hat\phi}{\partial x^0}.
\ee
This shows that the field operator $\hat\phi(x)$ satisfies Heisenberg's equations of motion.
It also satisfies the non-canonical commutation relations
\be
\left[\hat\phi(x),\hat\phi(y)\right]_-=-2i\sin\left(k_\mu(x-y)^\mu\right).
\label{scalar:cr}
\ee
The traditional objection against this kind of commutation relations,
with a r.h.s.~which does not vanish when $x\not=y$, is that they lead to a
theory with non-local interactions.
However note that the integral over the wave vector $\kk$ still has to be done.

\subsection{Non-coherent states}

It is now obvious to describe a general quantum field by an arbitrary
wave vector-dependent wave function $\psi_\kk$.
It determines an energy $E$ and a classical field $\phi(x)$ by
\be
E&=&\lu^3\int\upd\kk\,\langle \psi_\kk|\hat H|\psi_\kk\rangle,
\label{quant:energ3}\\
\phi(x)&=&\int\upd\kk\,\frac{\lu^{5/2}}{N(\kk)}\langle \psi_\kk|\hat\phi(x)|\psi_\kk\rangle.
\ee
Using the definition of $\hat \phi(x)$ the latter becomes
\be
\phi(x)&=&2\Re \int\upd\kk \,
\frac{\lu^{5/2}}{N(\kk)}e^{-ik_\mu x^\mu}
\langle \psi_\kk|\hat a|\psi_\kk\rangle
\label{quant:exp4}
\ee
Note that $\phi(x)$ is still a solution of the wave equation.
But now the wave function $\psi_\kk$ is not necessarily a coherent state.
As a consequence, the same classical field $\phi(x)$ can be obtained
both with and without involving coherent states.
However, in the case of non-coherent states part of the energy is of a quantum nature.
Indeed, the total energy, as given by (\ref {quant:energ3}), equals
\be
E&=&\hbar c\lu^3\int\upd\kk \,|\kk|\langle\psi_\kk|\hat a^\dagger\hat a\psi_\kk\rangle,
\label{quant:totenerg}
\ee
while the classical energy of the field (\ref {quant:exp4}) equals
(see (\ref{weq:classen}))
\be
\Ecl
&=&\hbar c\lu^3\int\upd\kk \,|\kk||\langle\psi_\kk|\hat a\psi_\kk\rangle|^2.
\ee
Now $E\ge \Ecl$ follows from the Cauchy-Schwarz inequality.

\subsection{Quantum correlations}
\label{sect:scalar:corr}

It would be appropriate to label the field operators with the wave vector $\kk$:
thus $\hat\phi_\kk(x)$ instead of $\hat\phi(x)$. 
This label is omitted when doing so does not create any ambiguity.
However, in the present section field operators involving different $\kk$-vectors appear.
Hence, the label is needed here.

So-called positive frequency and negative frequency parts of the field operator
are defined by
\be
\hat\phi^{(+)}_\kk(x)=e^{-ik_\mu x^\mu} \hat a
\quad\mbox{and}\quad
\hat\phi^{(-)}_\kk(x)=e^{ik_\mu x^\mu} \hat a^\dagger.
\ee
They are used to define a two-point function
\be
G(x,y)&=&
\int\upd\kk \,\frac{\lu^{5/2}}{N(\kk)}
\int\upd\kk' \,\frac{\lu^{5/2}}{N(\kk')}
\langle\psi_\kk|\hat\phi^{(-)}_\kk(x)\hat\phi^{(+)}_{\kk'}(y)\psi_{\kk'}\rangle.\cr
& &
\ee
Consider now the energy density $e(x)$ of the field at spacetime position $x$.
It is defined by
\be
e(x)
&=&
\frac{\hbar c}{\lu^2}\frac{\partial\,}{\partial x^0}\frac{\partial\,}{\partial y^0}
G(x,y)\bigg|_{y=x}
+\frac{\hbar c}{\lu^2}\sum_\alpha\frac{\partial\,}{\partial x^\alpha}\frac{\partial\,}{\partial y^\alpha}
G(x,y)\bigg|_{y=x}\cr
&=&
\frac{\hbar c\lu^3}{2(2\pi)^3}
\int\upd\kk \,\int\upd\kk' \,\frac{|\kk||\kk'|+\kk\cdot\kk'}{\sqrt{|\kk|\,|\kk'|}}
\langle\psi_\kk|\hat\phi^{(-)}_\kk(x)\hat\phi^{(+)}_{\kk'}(x)\psi_{\kk'}\rangle.
\label{energ:dens}
\ee
With this definition the integral of the energy density equals the total energy $E$ as given by  (\ref {quant:totenerg}).
Indeed,
\be
\int\upd\xx\,e(x)
&=&
\frac{\hbar c\lu^3}{2}
\int\upd\kk \,\int\upd\kk' \,\frac{|\kk||\kk'|+\kk\cdot\kk'}{\sqrt{|\kk|\,|\kk'|}}\delta(\kk-\kk')
\langle\psi_\kk|\hat a^\dagger\hat a\psi_{\kk'}\rangle\cr
&=&
\lu^3\int\upd\kk\,\hbar c|\kk|\langle\psi_{\kk}|a^\dagger a\psi_\kk\rangle=E.
\label{corr:intenenerg}
\ee

\subsection{Lorentz covariance}
\label{sect:scal:cov}

Given a Lorentz transformation $\Lambda$ the classical field transforms as
\be
\phi'(x)=\phi(\Lambda^{-1}x).
\ee
If $\Lambda$ is a spatial rotation $R$ then one can write
\be
\phi'(x)&=&2\Re \int\upd\kk \,
\frac{\lu^{5/2}}{N(\kk)}e^{-ik_\mu (\Lambda^\dagger x)^\mu}
\langle \psi_\kk|\hat a\psi_\kk\rangle\cr
&=&2\Re \int\upd\kk \,
\frac{\lu^{5/2}}{N(\kk)}e^{-i(\Lambda k)_\mu  x^\mu}
\langle \psi_\kk|\hat a\psi_\kk\rangle\cr
&=&2\Re \int\upd\kk' \,
\frac{\lu^{5/2}}{N(\kk')}e^{-ik'_\mu  x^\mu}
\langle \psi_{R^\dagger \kk'}|\hat a\psi_{R^\dagger \kk'}\rangle\cr
&=&\langle \psi'|\hat\phi(x)\psi'\ranglestar.
\ee
with $\psi'_\kk=\psi_{R^\dagger\kk}$.
Hence the rotation can be transferred from space-time
to the space of $\kk$-labeled wave functions.

Next consider a Lorentz boost of the form
\be
\Lambda=\left(\begin{array}{lccr}
\cosh\chi &0&0&\sinh\chi\\
0 &1 &0 &0\\
0 &0 &1 &0\\
\sinh\chi &0 &0 &\cosh\chi
 \end{array}\right).
\ee
Using
\be
\kk'=\left(\begin{array}{c}
            k^1\\
            k^2\\
            k^3\cosh \chi+|\kk|\sinh\chi 
           \end{array}\right),
           \label{scal:kprime}
\ee
one obtains
\be
\phi'(x)
&=&
2\Re \int\upd\kk \,
\frac{\lu^{5/2}}{N(\kk)}e^{-i k'_\mu  x^\mu}
\langle \psi_\kk|\hat a\psi_\kk\rangle.
\ee
Change the integration variable from $\kk$ to $\kk'$.
The Jacobian determinant equals $|\kk|/|\kk'|$. Hence one obtains
\be
\phi'(x)
&=&
2\Re \int\upd\kk' \,
\frac{\lu^{5/2}}{N(\kk')}e^{-i k'_\mu  x^\mu}
\langle \psi_\kk|\hat a\psi_\kk\rangle\cr
&=&2\Re\langle\psi'|\hat a\psi'\ranglestar,
\ee
with $\psi'_{\kk'}=\psi_\kk$.

This shows that also the Lorentz boost can be transferred from space-time
to the space of $\kk$-labeled wave functions.

\subsection{The quantum dimension}

This short section is inserted to support the paradigm shift from
quantum space as a deformation of classical space to quantum space as a
supplement orthogonal to classical space.
This section is not needed for the remainder of the paper.

The harmonic oscillator with Hamiltonian (\ref {scal:ham}) can be represented
as a one-dimensional mechanical oscillator, one for each wave vector $\kk$.
The easiest way to realize this is by adding a fifth dimension with coordinate
$x^4$ to the 4 dimensions of space-time. Given a coherent state $|z\ranglec$,
the quantum expectation of the position of this oscillator is $r\sqrt 2\Re z$.
Here, $r$ is a free parameter with the dimension of a length.
It is linked to the mass $m$  and the frequency
$c|\kk|$ of the oscillator by
\be
r=\sqrt{\frac{\hbar}{mc|\kk|}}.
\ee
For a classical wave $\phi(x)$ with complex profile function $f(\kk)$, defined by
(\ref {weq:exp}), the position of the oscillator is 
\be
x^4
&=&\sqrt 2\,r\lu^{-3/2} \Re f(\kk)e^{-i k_\mu  x^\mu}.
\ee
The penetration of the wave into the fifth dimension is proportional to the amplitude of the wave.
It oscillates with the frequency and the wave length of the wave.


\section{Electromagnetic fields}
\label{section:emf}

\subsection{The classical vector potential}

An electromagnetic wave traveling in direction 3 with electric
component in direction 1 can be described by the vector potential
\be
A(x)&\sim&
\left(
 \begin{array}{c}
 0\\1\\0\\0
 \end{array}\right)\cos(k(x^3-x^0)).
\ee
The electric and magnetic fields can be derived from the vector potential $A(x)$ by
 \be
 E_\alpha&=&-\frac{\partial A_\alpha}{\partial t}-c\frac{\partial A_0}{\partial x^\alpha},\cr
 B_\alpha&=&\sum_{\beta,\gamma}\epsilon_{\alpha,\beta,\gamma}\frac{\partial A_\gamma}{\partial x^\beta}.
 \ee
One then finds
\be
E_1&\sim& ck\sin(k(x^3-x^0)),
\ee
and $\displaystyle B_2=-\frac 1cE_1$ and  $E_2=E_3=cB_1=cB_3=0$.

Now let $\Xi(\kk)$ be a rotation matrix which rotates the arbitrary wave vector $\kk$
into the positive $z$-direction.
An explicit choice is found in the Appendix A.
Then an electromagnetic wave with wave vector $\kkph$ is described by the vector potential
with components $A_0(x)=0$ and
\be
A_\alpha(x)&\sim&\Re \Xi_{1,\alpha}(\kkph)e^{-i\kph_\mu x^\mu}.
\ee
After smearing out with a complex weight function $f(\kkph)$,
and inserting a normalization factor as before (see (\ref {weq:exp})), this becomes
\be
A_\alpha(x)=\Re 
\int\upd\kkph \,\frac{\lambda\lu}{N(\kkph)}f(\kkph)
\Xi_{1,\alpha}(\kkph)e^{-i\kph_\mu x^\mu}.
\ee
The parameter $\lambda$ could be absorbed into the weight function $f(\kkph)$
or could be combined with the unit of length $\lu$.
However,  it is kept for dimensional reasons.
It will be fixed later on.

The free electromagnetic wave has two possible polarizations.
The second linear polarization is obtained by replacing 
$\Xi_{1,\alpha}(\kkph)$ by $\Xi_{2,\alpha}(\kkph)$ in the previous expression.
In addition, the two polarizations can be combined by adding up the corresponding vector
potentials. The general expression is of the form
\be
A_\alpha(x)= \Re 
\int\upd\kkph \,\frac{\lambda\lu}{N(\kkph)}
\sum_{\beta=1,2}f_\beta(\kkph)
\Xi_{\beta,\alpha}(\kkph)e^{-i\kph_\mu x^\mu}.
\label{em:vp}
\ee

\subsection{Field operators}
\label{em:fieldop}

Because the electromagnetic wave has two polarizations it is obvious to consider a 2-dimensional
quantum harmonic oscillator instead of the single oscillator used in the section
on scalar bosons. 

\def\ah{\hat a_{\mbox{\tiny H}}}
\def\av{\hat a_{\mbox{\tiny V}}}
\def\ahdagger{\hat a^\dagger_{\mbox{\tiny H}}}
\def\avdagger{\hat a^\dagger_{\mbox{\tiny V}}}

Let $\ah$ and $\av$ be the annihilation operators for a photon with horizontal respectively
vertical polarization.
The Hamiltonian reads
\be
\hat H=\hbar c|\kkph|\left(\ahdagger\ah+\avdagger\av\right).
\label{em:ham}
\ee
Field operators are defined by
\be
\hat A_\alpha(x)&=&\frac 1{2}\lambda\varepsilon^{(H)}_\alpha(\kkph)
\left[e^{-i\kph_\mu x^\mu}\ah+e^{i\kph_\mu x^\mu}\ahdagger\right]\cr
& &+\frac 1{2}\lambda\varepsilon^{(V)}_\alpha(\kkph)
\left[e^{-i\kph_\mu x^\mu}\av+e^{i\kph_\mu x^\mu}\avdagger\right],
\label{em:potop}
\ee
with polarization vectors $\varepsilon^{(H)}_\alpha(\kkph)$ and $\varepsilon^{(V)}_\alpha(\kkph)$
given by two rows of the rotation matrix $\Xi$
\be
\varepsilon^{(H)}_\alpha(\kkph)
=\Xi_{1,\alpha}(\kkph)
\quad\mbox{ and }\quad
\varepsilon^{(V)}_\alpha(\kkph)
=\Xi_{2,\alpha}(\kkph).
\ee
Note that $\ah$ and $\av$ commute and that
$[\ah,\ahdagger]= \Io$ and $[\av,\avdagger]= \Io$.
This can be used to verify
that the field operators $\hat A_\alpha(x)$ satisfy Heisenberg's equation of motion.

Given a wave function $\psi_\kkph$ of the 2-dimensional harmonic oscillator the
expectation value of the field operators becomes
\be
A_\alpha(x)
&=&
\int\upd\kkph \frac{\lu^{5/2}}{N(\kkph)}\langle\psi_{\kkph}|\hat A_\alpha(x)\psi_{\kkph}\rangle\cr
&=&
\lambda \Re\int\upd\kkph \frac{\lu^{5/2}}{N(\kkph)}e^{-i\kph_\mu x^\mu}
\left(
\varepsilon^{(H)}_{\alpha}(\kkph)
\langle\psi_{\kkph}|\ah\psi_{\kkph}\rangle
+\varepsilon^{(V)}_{\alpha}(\kkph)
\langle\psi_{\kkph}|\av\psi_{\kkph}\rangle\right).
\label{photon:classfield}
\ee
This is of the form (\ref {em:vp}) with
\be
f_1(\kkph)=\lu^{3/2}\langle\psi_{\kkph}|\ah\psi_{\kkph}\rangle
\quad\mbox{ and }\quad
f_2(\kkph)=\lu^{3/2}\langle\psi_{\kkph}|\av\psi_{\kkph}\rangle.
\ee

Operator-valued electric and magnetic fields are defined by
 \be
 \hat E_\alpha&=&-\frac{\partial \hat A_\alpha}{\partial t},\cr
 \hat B_\alpha&=&\sum_{\beta,\gamma}\epsilon_{\alpha,\beta,\gamma}\frac{\partial \,}{\partial\xx_\beta}
\hat A_\gamma.
 \ee
Gauss's law in absence of charges is satisfied. Indeed, one has
\be
\sum_\alpha\frac{\partial\,}{\partial x^\alpha}\hat E_\alpha
&=&\frac 1{2} \lambda c|\kkph|\left(\sum_\alpha \kph_\alpha \varepsilon^{(H)}_\alpha(\kkph)\right)
\left[ e^{-i\kph_\mu x^\mu}\ah+e^{i\kph_\mu x^\mu}\ahdagger\right]
\cr
& &+\frac 1{2} \lambda c|\kkph|\left(\sum_\alpha \kph_\alpha \varepsilon^{(V)}_\alpha(\kkph)\right)
\left[ e^{-i\kph_\mu x^\mu}\av+e^{i\kph_\mu x^\mu}\avdagger\right]
\cr
&=&0,
\label{photon:gauss}
\ee
because
\be
\sum_\alpha \kph_\alpha \varepsilon^{(H)}_\alpha(\kkph)
=\left(\Xi(\kkph)\kkph\right)_1
=|\kkph|(\eev_3)_1
\label{em:hororth}
\ee
vanishes, as well as a similar expression for the vertical polarization.

Finally let us calculate the commutation relations
\be
\left[\hat A_\alpha(x),\hat A_\beta(y)\right]_-
&=&-\frac i{2}\lambda^2\left(
\varepsilon^{(H)}_\alpha(\kkph)\varepsilon^{(H)}_\beta(\kkph)
+\varepsilon^{(V}_\alpha(\kkph)\varepsilon^{(V)}_\beta(\kkph)\right)\sin( \kph_\mu(x-y)^\mu)\cr
&=&-\frac i{2}\lambda^2\left[
\Xi_{\alpha,1}(\kkph)\Xi_{\beta,1}(\kkph)
+\Xi_{\alpha,2}(\kkph)\Xi_{\beta,2}(\kkph)\right]
\sin (\kph_\mu(x-y)^\mu).\cr
& &
\ee
These commutation relations differ from the standard ones
in the first place because the integration over the $\kkph$-vector still has to be
carried through. 

\subsection{Lorentz transformations -- spatial rotations}

\def\bh{\hat b_{\mbox{\tiny H}}}
\def\bv{\hat b_{\mbox{\tiny V}}}

The electromagnetic vector potential has been studied above in the temporal gauge,
which is characterized by the vanishing of the zeroth component $A_0(x)$. 
The vector potential transforms as a 4-vector. In particular, this means that the temporal
gauge will be broken by a Lorentz boost. This complicates the study of general
Lorentz transformations. 
Note that the Lorentz covariance has been checked in Section \ref {sect:scal:cov}.
What follows is therefore restricted to spatial rotations.

Consider a spatial rotation matrix $\Lambda$. 
The temporal gauge is not broken.
The new vector potential $A'$ relates to the old one by
\be
A'(\Lambda x)=\Lambda A(x).
\ee
Let $R$ denote the 3-by-3 rotational submatrix of $\Lambda$.
One calculates
\be
A'_\alpha(x)
&=&\sum_\beta R_{\alpha,\beta}A_\beta(\Lambda^\dagger x)\cr
&=&
\lambda\int\upd\kkph \frac{\lu^{5/2}}{N(\kkph)}
\sum_\beta R_{\alpha,\beta}\varepsilon^{(H)}_{\beta}(\kkph)
\Re \langle\psi_{\kkph}|\ah\psi_{\kkph}\rangle
e^{-i(\Lambda k)_\mu x^\mu}
\cr& &
+
\lambda\int\upd\kkph \frac{\lu^{5/2}}{N(\kkph)}
\sum_\beta R_{\alpha,\beta}\varepsilon^{(V)}_{\beta}(\kkph)
\Re \langle\psi_{\kkph}|\av\psi_{\kkph}\rangle
e^{-i(\Lambda k)_\mu x^\mu}.
\ee
Note that one can write
\be
\Xi(R \kkph)R=M_\Lambda(\kkph)\Xi(\kkph),
\label{em:mlambda}
\ee
where $M_\Lambda(\kkph)$ is a rotation around the third axis.
It brings the polarization axes into their conventional position.
It implies
\be
\sum_\beta R_{\alpha,\beta}\varepsilon^{(H)}_{\beta}(\kkph)
&=&\sum_\beta \Xi_{\beta,\alpha}(R\kkph)[M_\Lambda(\kkph)]_{\beta,1}
\ee
and
\be
\sum_\beta R_{\alpha,\beta}\varepsilon^{(V)}_{\beta}(\kkph)
&=&\sum_\beta \Xi_{\beta,\alpha}(R\kkph)[M_\Lambda(\kkph)]_{\beta,2}
\ee
One has therefore
\be
A'_\alpha(x)
&=&\lambda
\int\upd\kkph \frac{\lu^{5/2}}{N(\kkph)}
\sum_\beta \Xi_{\beta,\alpha}(R\kkph)[M_\Lambda(\kkph)]_{\beta,1}
\Re\langle\psi_{\kkph}|\ah\psi_{\kkph}\rangle
e^{-i(\Lambda k)_\mu x^\mu}\cr
& &+\lambda
\int\upd\kkph \frac{\lu^{5/2}}{N(\kkph)}
\sum_\beta \Xi_{\beta,\alpha}(R\kkph)[M_\Lambda(\kkph)]_{\beta,2}
\Re\langle\psi_{\kkph}|\av\psi_{\kkph}\rangle
e^{-i(\Lambda k)_\mu x^\mu}\cr
&=&
\lambda
\int\upd\kkph \frac{\lu^{5/2}}{N(\kkph)}
\varepsilon^{(H)}_{\alpha}(R\kkph)
\Re\langle\psi_{\kkph}|\bh\psi_{\kkph}\rangle
e^{-i(\Lambda k)_\mu x^\mu}\cr
& &+\lambda
\int\upd\kkph \frac{\lu^{5/2}}{N(\kkph)}
\varepsilon^{(V)}_{\alpha}(R\kkph)
\Re\langle\psi_{\kkph}|\bv\psi_{\kkph}\rangle
e^{-i(\Lambda k)_\mu x^\mu}
\ee
with
\be
\bh&=&[M_\Lambda(\kkph)]_{1,1}\ah+[M_\Lambda(\kkph)]_{1,2}\av,\\
\bv&=&[M_\Lambda(\kkph)]_{2,1}\ah+[M_\Lambda(\kkph)]_{2,2}\av.
\ee
A change of integration variables $\kkph'=R\kkph$ then gives
\be
A'_\alpha(x)
&=&\lambda
\int\upd\kkph' \frac{\lu^{5/2}}{N(\kkph')}
\varepsilon^{(H)}_{\alpha}(\kkph')
\Re\langle\psi_{\kkph}|\bh\psi_{\kkph}\rangle
e^{-ik'_\mu x^\mu}\cr
& &+\lambda
\int\upd\kkph' \frac{\lu^{5/2}}{N(\kkph')}
\varepsilon^{(V)}_{\alpha}(\kkph')
\Re\langle\psi_{\kkph}|\bv\psi_{\kkph}\rangle
e^{-ik'_\mu x^\mu}.
\ee
The effect of the rotation $\Lambda$ is a rotation of the $\kkph$-vector together with a
transformation of the annihilation operators.
The latter is a symmetry of the two-dimensional harmonic oscillator.
Note that $\bh$ and $\bv$ are again annihilation operators. 

The transformation of the annihilation operators
can be transferred to a rotation in the Hilbert space of wave functions.
Indeed, let $|m,n\rangle$ be the eigenstate satisfying
\be
\ah|m,n\rangle=m|m-1,n\rangle
\quad\mbox{and}\quad
\av|m,n\rangle=n|m,n-1\rangle,
\ee
and let $|m,n\rangle'$ be the eigenstate satisfying
\be
\bh|m,n\rangle'=m|m-1,n\rangle'
\quad\mbox{and}\quad
\bv|m,n\rangle'=n|m,n-1\rangle'.
\ee
Then a unitary operator $U_\Lambda(\kkph)$ is defined by
\be
U_\Lambda(\kkph)|m,n\rangle=|m,n\rangle'.
\ee
It satisfies
\be
U_\Lambda(\kkph)\ah U^\dagger_\Lambda(\kkph)=\bh
\quad\mbox{and}\quad
U_\Lambda(\kkph)\av U^\dagger_\Lambda(\kkph)=\bv.
\ee
Hence the previous result becomes
\be
A'_\alpha(x)
&=&\lambda
\int\upd\kkph' \frac{\lu^{5/2}}{N(\kkph')}
\varepsilon^{(H)}_{\alpha}(\kkph')
\Re\langle\psi'_{\kkph'}|\ah\psi'_{\kkph'}\rangle
e^{-ik'_\mu x^\mu}\cr
& &+\lambda
\int\upd\kkph' \frac{\lu^{5/2}}{N(\kkph')}
\varepsilon^{(V)}_{\alpha}(\kkph')
\Re\langle \psi'_{\kkph'}|\av\psi'_{\kkph'}\rangle
e^{-ik'_\mu x^\mu}\cr
&=&\lambda\Re
\int\upd\kkph \frac{\lu^{5/2}}{N(\kkph)}
\varepsilon^{(H)}_{\alpha}(\kkph)
\langle\psi'_{\kkph}|\ah\psi'_{\kkph}\rangle
e^{-i\kph_\mu x^\mu}\cr
& &+\lambda\Re
\int\upd\kkph \frac{\lu^{5/2}}{N(\kkph)}
\varepsilon^{(V)}_{\alpha}(\kkph)
\langle\psi'_{\kkph}|\av\psi'_{\kkph}\rangle
e^{-i\kph_\mu x^\mu}
\ee
with
\be
\psi'_{\kkph'}=U_\Lambda(\kkph)\psi_\kkph.
\ee
This shows that a rotation of the frame of reference can be represented
by a rotation of the $\kkph$-vector
together with a $\kkph$-dependent unitary transformation of the space of wave functions. 

\subsection{Quanta}
\label{subsect:quanta}

Given two complex numbers $z$ and $w$ let $|z,w\ranglec$ denote the product
of the two coherent wave functions $|z\ranglec$ and $|w\ranglec$.
One has
$\ah |z,w\ranglec=z|z,w\ranglec$ and $\av|z,w\ranglec=w|z,w\ranglec$.
The quantum expectation of the Hamiltonian (\ref {em:ham}) 
with $\psi_\kkph=|F(\kkph),0\ranglec$ equals (see also (\ref {quant:totenerg}))
\be
E
&=&\hbar c\lu^3
\int\upd\kkph\, |\kkph|
\langle\psi_{\kkph}|\ahdagger\ah\psi_{\kkph}\rangle\cr
&=&\hbar c\lu^3
\int\upd\kkph \, |\kkph|
|F(\kkph)|^2\cr
&=&
\hbar c
\int\upd\kkph \, |\kkph|
|f(\kkph)|^2
\label{photon:quante}
\ee
with $f(\kkph)=\lu^{3/2}F(\kkph)$.

On the other hand follows from (\ref {photon:classfield}) that
\be
E_\alpha(x)
&=&-c\frac{\partial A_\alpha}{\partial x^0}\cr
&=&\lambda c\int\upd\kkph \,\frac{\lu^{5/2}}{N(\kkph)}|\kkph|
\varepsilon^{(H)}_{\alpha}(\kkph)
\Re iF(\kkph)e^{-i\kph_\mu x^\mu}
\ee
and
\be
B_\alpha(x)
&=&\lambda
\sum_{\beta,\gamma}\epsilon_{\alpha,\beta,\gamma}
\int\upd\kkph \,\frac{\lu^{5/2}}{N(\kkph)}\kph_\beta
\varepsilon^{(H)}_{\gamma}(\kkph)
\Re iF(\kkph)e^{-i\kph_\mu x^\mu}.
\ee
These expressions can be used to calculate the total energy $\Ecl$ of the electromagnetic wave.
Denote $\varepsilon_0$ the permittivity of the vacuum.
One obtains
\be
\Ecl
&=&\frac 12\varepsilon_0\int\upd \xx\,\sum_\alpha(E_\alpha^2+c^2B_\alpha^2)\cr
&=&-\frac 12\varepsilon_0\lambda^2c^2\lu^5
\int\upd\kkph \,\frac{1}{N(\kkph)}\int\upd\kkph' \,\frac{1}{N(\kkph')}\cr
& &\times
\left[|\kkph|\,|\kkph'|\sum_\alpha\varepsilon^{(H)}_{\alpha}(\kkph)\varepsilon^{(H)}_{\alpha}(\kkph')
+\sum_\alpha\sum_{\beta,\gamma}\sum_{\beta',\gamma'}\epsilon_{\alpha,\beta,\gamma}\epsilon_{\alpha,\beta',\gamma'}
\kph_\beta k'_{\beta'}
\varepsilon^{(H)}_{\gamma}(\kkph)\varepsilon^{(H)}_{\gamma'}(\kkph')
\right]\cr
& &\times
\int\upd \xx\,\frac 14
\left(F(\kkph)e^{-i\kph_\mu x^\mu}-\overline{F(\kkph)}e^{i\kph_\mu x^\mu}\right)
\left(F(\kkph')e^{-ik'_\mu x^\mu}-\overline{F(\kkph')}e^{ik'_\mu x^\mu}\right).
\ee
The integral over $x$ evaluates to
\be
-\frac 12(2\pi)^3\delta(\kkph-\kkph')|F(\kkph)|^2+
\frac 14(2\pi)^3\delta(\kkph+\kkph')\left[F(\kkph)F(-\kkph)+\overline{F(\kkph)F(-\kkph)}\right].
\label{photon:vanish}
\ee
The latter term does not contribute. See the Appendix \ref {appendix:vanish}.
Therefore the total energy becomes
\be
\Ecl&=&\frac 1{8}\varepsilon_0\lambda^2c^2\lu^5
\int\upd\kkph \,\frac{1}{|\kkph|}|F(\kkph)|^2
\cr& &\times
\left[|\kkph|^2\sum_\alpha[\varepsilon^{(H)}_{\alpha}(\kkph)]^2
+\sum_\alpha\sum_{\beta,\gamma}\sum_{\beta',\gamma'}\epsilon_{\alpha,\beta,\gamma}\epsilon_{\alpha,\beta',\gamma'}
\kph_\beta \kph_{\beta'}
\varepsilon^{(H)}_{\gamma}(\kkph)\varepsilon^{(H)}_{\gamma'}(\kkph)
\right].\cr
& &
\ee
Because $\Xi(\kkph)$ is a rotation matrix one has
\be
\sum_\alpha[\varepsilon^{(H)}_{\alpha}(\kkph)]^2=\sum_\alpha \Xi_{1,\alpha}(\kkph)^2=1
\ee
and
\be
\sum_\alpha\sum_{\beta,\gamma}\sum_{\beta',\gamma'}
\epsilon_{\alpha,\beta,\gamma}
\epsilon_{\alpha,\beta',\gamma'}\kph_\beta \kph_{\beta'}
\Xi_{1,\gamma}(\kkph)\Xi_{1,\gamma'}(\kkph)
&=&|\kkph\times \Xi_1|^2\cr
&=&|\kkph|^2|\Xi_1|^2-\kkph\cdot\Xi_1(\kkph)\cr
&=&|\kkph|^2-[\Xi(\kkph)\kkph]_1.
\ee
Remember that $\Xi(\kkph)\kkph$ points into the third direction.
Hence, its first component vanishes. The above expression simplifies to
\be
\Ecl&=&\frac 1{4}\varepsilon_0\lambda^2c^2\lu^5\int\upd\kkph \,|\kkph||F(\kkph)|^2\cr
&=&\frac 1{4}\varepsilon_0\lambda^2c^2\lu^2\int\upd\kkph \,|\kkph||f(\kkph)|^2
\ee
This leads to the conclusion that the classical energy of the coherent electromagnetic wave coincides
with the quantum expectation (\ref {photon:quante}) of the Hamiltonian  (\ref {em:ham})
provided that the free parameter $\lambda\lu$ has the value
\be
\lambda\lu=4\sqrt{\frac{\hbar}{\varepsilon_0 c}}=4\frac{\hbar}{\uc}\sqrt{\pi\alpha},
\label{ef:nc}
\ee
where $\uc$ is the unit of charge and $\alpha=\uc^2/4\pi\varepsilon_0\hbar c$ is the fine structure constant.

\subsection{Stokes operators}

The Stokes operators are proportional to the the angular momentum operators of the photon.
They are given by
\be
\hat S_0&=&\ahdagger\ah+\avdagger\av,\\
\hat S_1&=&\ahdagger\av+\avdagger\ah,\\
\hat S_2&=&i\avdagger\ah-i\ahdagger\av,\\
\hat S_3&=&\ahdagger\ah-\avdagger\av.
\ee
In particular, $-S_2$ is (proportional to) the spin of the photon in the direction of motion $\kkph$.
See for instance \cite{JR76}, Section 2.8.

\def\fh{f^{\mbox{\tiny H}}}
\def\fv{f^{\mbox{\tiny V}}}
\def\Fh{F^{\mbox{\tiny H}}}
\def\Fv{F^{\mbox{\tiny V}}}

Take for instance a product of coherent states $\psi_\kkph=|\Fh(\kkph),\Fv(\kkph)\ranglec$.
Then, using
\be
S_i&=&\lu^{3}\int\upd\kkph\,\langle \Fh(\kkph),\Fv(\kkph)|\hat S_i|\Fh(\kkph),\Fv(\kkph)\ranglec,
\ee
one finds
\be
S_0
&=&\int\upd\kkph \,\left(|\fh|^2+|\fv|^2\right)\\
S_1
&=&\int\upd\kkph \,\left(\overline{\fv}\fh+\overline{\fh}\fv\right)\\
S_2
&=&i\int\upd\kkph \,\left(\overline{\fv}\fh-\overline{\fh}\fv\right)\\
S_3
&=&\int\upd\kkph \,\left(|\fh|^2-|\fv|^2\right).
\ee

\subsection{Single photon states}
\label{sect:singphot}

Consider $\kkph$-dependent superpositions of a polarized one-photon state and the vacuum
of the form 
\be
\psi_\kkph=\sqrt{\rho(\kkph)}e^{i\phi(\kkph)}|1,0\rangle+\sqrt{1-\rho(\kkph)}|0,0\rangle.
\ee
The energy of the electromagnetic wave equals
\be
E&=&\hbar c\int\upd\kkph\,|\kkph|\rho(\kkph).
\ee
The wave vector distribution $|\kkph|\rho(\kkph)$ must be integrable to keep the total energy finite.
In particular, $\rho(\kkph)$  cannot be taken constant.
Therefore, the superposition of the one-photon state and the vacuum is a necessity.

The quantum expectation of the vector potential evaluates to
\be
A_\alpha(x)
&=&
\lambda\int\upd\kkph\, \frac{\lu^{5/2}}{N(\kkph)}\sqrt{\rho(\kkph)(1-\rho(\kkph))}
\varepsilon^{(H)}_{\alpha}(\kkph)
\Re e^{i\phi(\kkph)}e^{-i\kph_\mu x^\mu}.\cr
& &
\ee
Note that the contribution to the classical electromagnetic field comes from the region where
the overlap with the vacuum is neither 0 nor 1.

The one-photon state discussed above is linearly polarized.
Indeed, one verifies immediately that $S_2=0$.
On the other hand, an example of circularly polarized one-photon states is obtained by
choosing
\be
\psi_\kkph=
\sqrt{\rho(\kkph)}e^{i\phi(\kkph)}
\frac{1}{\sqrt 2}(|1,0\rangle\pm i|0,1\rangle)
+\sqrt{1-\rho(\kkph)}|0,0\rangle.
\ee
The spin is given by
$\displaystyle
S_2=\pm \lu^{3}\int\upd\kkph\,\rho(\kkph)
$.
Note that
$\displaystyle
\frac{1}{\sqrt 2}(|1,0\rangle\pm i|0,1\rangle)
$
is a wave function of the 2-dimensional harmonic oscillator, just like $|1,0\rangle$
or $|0,1\rangle$. Both linearly and circularly polarized one-photon states exists
in the present theory.


\section{Conclusions}

The present paper shows that one can introduce in QED wave functions $\psi_\kkph$,
which describe a free photon field,
and which are normalized in such a way that $||\psi_\kkph||=1$
for each individual wave vector $\kkph$.
The wave functions are made dimensionless by introduction of a
physical constant, which is the product $\lambda\lu$ in my notations. 
It is fixed by requiring that the quantum expectation of the energy
of an electromagnetic wave is always
larger than or equal to the classical energy of the corresponding
classical wave, with equality when the wave is coherent.
The resulting value can be expressed in existing physical constants --- see (\ref {ef:nc}).

The introduction of these wave functions, which are normalized in an unconventional manner,
leads to further modifications of the standard theory of QED.
The Hamiltonian of the free photon fields is that of a 2-dimensional harmonic oscillator
Its frequency $c|\kkph|$, because of its dependence on the wave vector $\kkph$,
can be seen as a multiplication operator acting on the wave vector dependence of the wave functions.
Such dependence is present also in standard QED. However, integrations over the $\kkph$-vector
now move from the operator expressions towards the evaluation of expectation values.

The Lorentz covariance of the new theory has been verified explicitly in the case of 
the scalar boson field. It is omitted for the electromagnetic fields because of
the complexity of the resulting expressions. Indeed, 
all possible relativistic effects such as red shift appear in their most general form.
They are of course of interest but fall out of the scope of the present paper.

To illustrate the present approach single photon wave functions are considered in Section 
\ref {sect:singphot}. Idealized monochromatic photons have vanishing electromagnetic field,
as it should be. However, realistic wave functions coincide with the wave function
of the vacuum state for large values of the wave vector. 
In the overlap region classical electromagnetic waves do appear.
An open question is whether experimentally measured electromagnetic fields
produced by single photons can be matched with wave functions of the type introduced here.

In Part II of the present paper the methodology developed in this first part
is extended to the fermionic case.
Part III deals with interactions between photons and electrons.
The main argument for splitting the manuscript into 3 parts is that 
the interaction part contains many issues where difficult choices have to be made,
while Parts I and II are self-contained and rather straightforward.

\appendix 
\section*{Appendices}

\section{Polarization of electromagnetic waves}
\label{appendix:pol}

Let us first make an explicit choice for the rotation matrix $\Xi(\kk)$.
By definition it rotates the vector $\kk$ into the positive third direction.
Choose to do this by rotating around the first axis and then around the second axis.
This gives a matrix of the form
\be
\Xi(\kk)
&=&
\left(\begin{array}{lcr}
          \cos\beta &0 &-\sin\beta\\
          0 &1&0\\
          \sin\beta &0 &\cos\beta
         \end{array}\right)
\left(\begin{array}{lcr}
          1 &0 &0\\
          0 &\cos\alpha&-\sin\alpha\\
          0 &\sin\alpha &\cos\alpha
         \end{array}\right)\cr
&=&
\left(\begin{array}{lcr}
          \cos\beta &-\sin\alpha\sin\beta &-\sin\beta\cos\alpha\\
          0 &\cos\alpha&-\sin\alpha\\
          \sin\beta &\sin\alpha\cos\beta &\cos\alpha\cos\beta
         \end{array}\right).
\ee
The first rotation eliminates component 2, the second eliminates component 1.
From these requirements one deduces that
\be
\cos\alpha=\frac{k_3}{\sqrt{k_2^2+k_3^2}},
\qquad
\sin\alpha=\frac{k_2}{\sqrt{k_2^2+k_3^2}},
\ee
and
\be
\cos\beta=\frac{1}{|\kk|}\sqrt{k_2^2+k_3^3},
\qquad
\sin\beta=\frac{k_1}{|\kk|}.
\ee
It is now straightforward to verify that $\Xi(\kk)\kk=|\kk|e_3$.

Next let $R$ be an arbitrary rotation matrix and $\Lambda$ the corresponding
4-dimensional Lorentz transformation. We want to calculate
\be
M_\Lambda(\kk)=\Xi(R\kk)R\Xi(\kk)^\dagger.
\ee
We know that $M_\Lambda(\kk)$ is of the form
\be
M_\Lambda(\kk)&=&
\left(\begin{array}{lcr}
          \cos\gamma &-\sin\gamma &0\\
          \sin\gamma &\cos\gamma&0\\
          0 &0&1\\
         \end{array}\right).
\ee
Use the notation $\kk'=R\kk$, and $\alpha'$, $\beta'$ for the value of
$\alpha$ respectively $\beta$ when evaluated at $\kk'$ instead of $\kk$.
Then the first row of the matrix $\Xi(\kk')^\dagger M_\Lambda(\kk)$
evaluates to
\be
\big[
\cos\beta'\cos\gamma,\quad
-\cos\beta'\sin\gamma,
\quad\sin\beta'
\big].
\ee
The first row of the matrix $R\Xi(\kk)^\dagger$ evaluates to
\be
&&\big[
R_{1,1}\cos\beta-(R_{1,2}\sin\alpha+R_{1,3}\cos\alpha)\sin\beta,\quad
R_{1,2}\cos\alpha-R_{1,3}\sin\alpha,\cr
&&\quad
R_{1,1}\sin\beta+(R_{1,2}\sin\alpha+R_{1,3}\cos\alpha)\cos\beta
\big].
\ee
Equating both rows yields
\be
\cos\beta'\cos\gamma&=&R_{1,1}\cos\beta-(R_{1,2}\sin\alpha+R_{1,3}\cos\alpha)\sin\beta,\\
-\cos\beta'\sin\gamma&=&R_{1,2}\cos\alpha-R_{1,3}\sin\alpha,\\
\sin\beta'&=&R_{1,1}\sin\beta+(R_{1,2}\sin\alpha+R_{1,3}\cos\alpha)\cos\beta.
\ee
The latter implies
\be
\cos\gamma&=&\frac{R_{1,1}-\sin\beta\sin\beta'}{\cos\beta\cos\beta'},\\
\sin\gamma&=&-\frac{R_{1,2}\cos\alpha-R_{1,3}\sin\alpha}{\cos\beta'}.
\ee

\section{Vanishing contribution}
\label {appendix:vanish}
Here it is shown that (\ref {photon:vanish}) vanishes.
One calculates
\be
& &
\frac 1{2\lu^2}\lambda^2\int\upd\kkph \,\int\upd\kkph' \,
\frac{1}{\sqrt{\omega(\kkph)\omega(\kkph')}}f(\kkph)f(\kkph')\cr
& &\times
\left[|\kkph|\,|\kkph'|\sum_\alpha \Xi_{1,\alpha}(\kkph)\Xi_{1,\alpha}(\kkph')
+\sum_\alpha\sum_{\beta,\gamma}\epsilon_{\alpha,\beta,\gamma}
\sum_{\beta',\gamma'}\epsilon_{\alpha,\beta',\gamma'}\kph_\beta k'_{\beta'}
\Xi_{1,\gamma}(\kkph)\Xi_{1,\gamma'}(\kkph')
\right]\cr
& &\times
\frac{1}{2\pi^3}\int\upd \xx\,\sin(|\kkph|x_0+\kkph\cdot\xx)\sin(|\kkph'|x_0+\kkph'\cdot\xx).
\ee
A further calculation gives
\be
& &
\frac{1}{2\pi^3}\int\upd \xx\,\sin(|\kkph|x_0+\kkph\cdot\xx)\sin(|\kkph'|x_0+\kkph'\cdot\xx)\cr
&=&\frac 12\delta(\kkph-\kkph')-\frac 12\delta(\kkph+\kkph')\cos(2|\kkph|x_0).\cr
& &
\ee
Hence the previous expression reduces to
\be
E&=&
\frac 1{4\lu^2}\lambda^2c^2\int\upd\kkph \,\frac{1}{c|\kkph|}f(\kkph)^2\cr
& &\times
\left[|\kkph|^2\sum_\alpha \Xi_{1,\alpha}(\kkph)^2
+\sum_\alpha\sum_{\beta,\gamma}\sum_{\beta',\gamma'}\epsilon_{\alpha,\beta,\gamma}
\epsilon_{\alpha,\beta',\gamma'}\kph_\beta \kph_{\beta'}
\Xi_{1,\gamma}(\kkph)\Xi_{,\gamma'}(\kkph)
\right]\cr
& &-\frac 1{4\lu^2}\lambda^2c^2\int\upd\kkph \,\frac{1}{c|\kkph|}f(\kkph)f(-\kkph)\cos(2|\kkph|x_0)\cr
& &\times
\left[|\kkph|^2\sum_\alpha \Xi_{1,\alpha}(\kkph)\Xi_{1,\alpha}(-\kkph)
-\sum_\alpha\sum_{\beta,\gamma}\sum_{\beta',\gamma'}
\epsilon_{\alpha,\beta,\gamma}\epsilon_{\alpha,\beta',\gamma'}\kph_\beta \kph_{\beta'}
\Xi_{1,\gamma}(\kkph)\Xi_{1,\gamma'}(-\kkph)
\right].\cr
& &
\label {photon:energy}
\ee
Because $\Xi(\kkph)$ is a rotation matrix one has
\be
\sum_\alpha \Xi_{1,\alpha}(\kkph)^2&=&1
\ee
and
\be
\sum_\alpha\sum_{\beta,\gamma}\sum_{\beta',\gamma'}
\epsilon_{\alpha,\beta,\gamma}
\epsilon_{\alpha,\beta',\gamma'}\kph_\beta \kph_{\beta'}
\Xi_{1,\gamma}(\kkph)\Xi_{1,\gamma'}(\kkph')
&=&|\kkph\times \Xi_1|^2\cr
&=&|\kkph|^2|\Xi_1|^2-\kkph\cdot\Xi_1(\kkph)\cr
&=&|\kkph|^2-[\Xi(\kkph)\kkph]_1.
\ee
Remember that $\Xi(\kkph)\kkph$ points into the third direction.
Hence, its first component vanishes. The first contribution to the energy becomes
\be
E=\frac 1{2\lu^2}\lambda^2 c\int\upd\kkph \,|\kkph|f(\kkph)^2.
\ee
The second contribution vanishes because
\be
& &\sum_\alpha\sum_{\beta,\gamma}\sum_{\beta',\gamma'}
\epsilon_{\alpha,\beta,\gamma}\epsilon_{\alpha,\beta',\gamma'}\kph_\beta \kph_{\beta'}
\Xi_{1,\gamma}(\kkph)\Xi_{1,\gamma'}(-\kkph)\cr
&=&[\kkph\times\Xi_1(\kkph)]\cdot [\kkph\times\Xi_1(-\kkph)]\cr
&=&|\kkph|^2\Xi_1(\kkph)\cdot \Xi_1(-\kkph)
-[\kkph\cdot \Xi_1(-\kkph)][\Xi_1\cdot \kkph]\cr
&=&|\kkph|^2\Xi_1(\kkph)\cdot \Xi_1(-\kkph)\cr
&=&|\kkph|^2\sum_\alpha \Xi_{1,\alpha}(\kkph)\Xi_{1,\alpha}(-\kkph).
\ee
Hence the two contributions of the second term of (\ref{photon:energy}) cancel each other.


\end{document}